\documentclass[12pt]{article}
\input epsf.sty

\begin{document}
\begin{center}

{\Large \textbf{Superflares of H$_2$O Maser Emission\\
\vspace{3mm}
in the Protostellar Object IRAS\,18316$-$0602}}
\bigskip

{\large {\bf\copyright 2017  E.E. Lekht$^1$, M.I. Pashchenko$^1$,
G.M. Rudnitskij$^1$, \protect\\
and A.M. Tolmachev$^2$} }

\medskip

{\it $^1$M.V.~Lomonosov Moscow State University,
Sternberg Astronomical Institute, \protect\\
13 Universitetskii prospekt, Moscow, 119234 Russia

$^2$Pushchino Radio Astronomy Observatory, Astrospace Center \protect\\
of the Lebedev Institute of Physics,
Russian Academy of Sciences, \protect\\
Pushchino, Moscow Region, 142290 Russia\\
{\rm e-mail:~~~lekht@sai.msu.ru~~~~eelekht@mail.ru} }
\end{center}

\bigskip


The results of the study of the maser emission source IRAS
18316$-$0602 in the H$_2$O line at $\lambda =1.35$~cm are
reported. The observations have been carried out at the RT-22
radio telescope of the Pushchino Radio Astronomy Observatory
(Russia) since June 2002 until March 2017. Three superflares
have been detected, in 2002, 2010, and 2016, with peak flux
densities of 3400, 19,000, and 46,000 Jy, respectively. The
results of the analysis of the superflares are given. The flares
took place during periods of high maser activity in a narrow
interval of radial velocities (40.5--42.5~km/s) and could be
associated with the passage of a strong shock wave. During our
monitoring the emission of three groups of features at radial
velocities of about 41, 42, and 43 km/s dominated. The flare of
2016 was accompanied by a considerable increase in the flux
densities of several features in an interval of 35--56~km/s.

\bigskip

(IRAS\,18316$-$0602)

\newpage

\section{Introduction}

The source IRAS\,18316$-$0602 is located in a region of active star formation.
It is associated with the ultracompact HII region G25.65+1.05 (Kurtz et~al. [1];
Jenness et~al. [2]; Walsh et~al. [3]) and a molecular outflow (McCutcheon et~al. [4]). The radio source coincides spatially with an unresolved IR source and submillimeter emission at 350, 450, and 850 $\mu$m (Hunter et~al. [5]; Walsh et~al. [6]). Shepherd and Churchwell [7] showed that the CO bipolar outflow is centered on the radio source. S\'anchez-Monge et~al. [8] mapped the molecular outflow in the SiO $J=2-1$ line and found that the redshifted ($+48.5<V_{\textrm{red}}<88.1$~km/s) and blueshifted ($+5.9<V_{\textrm{blue}}<+39.5$~km/s) maxima are offset in declination by $40^{\prime\prime}$.

The adopted distance to the source is kinematic. Molinari et~al. [9] give 3.17~kpc, and Sunada et~al. [10], 2.7~kpc. The value 3.17~kpc is considered as preferable.

Palla et~al. [11] observed toward IRAS\,18316$-$0602 strong maser emission in the H$_2$O at a radial velocity of 45.17~km/s with a peak flux density of 725~Jy. In July 1994 it was 109~Jy (Jenness et~al. [2]). In 1995 the H$_2$O emission was observed in a broad interval of radial velocities; the main emission took place at 41.30 and 45.26~km/s with peak flux densities of 452 and 260~Jy [12].

OH maser emission in the main lines 1665 and 1667~MHz was detected by Edris et~al. [13]. In the 1612-MHz satellite line emission is thermal, and the 1720-MHz line is in absorption. In addition, maser emission in methanol CH$_3$OH lines was observed; e.g., Walsh et~al. [3], Szymczak et~al. [14], Surcis et~al. [15]. It is associated with the radio source, probably, with the disk, but not with the bipolar outflow.

The positions of the ultracompact HII region, OH and H$_2$O, near- and far-infrared sources are sufficiently close (e.g., Jenness et~al. [2]).

\section{Observations and Data Presentation}

The maser source IRAS 18316$-$0602 was included into the program of our 1.35-cm water vapor line monitoring at the RT-22 radio telescope in Pushchino in 2002. The system noise temperature was 130--250~K depending on weather conditions. The half-power beamwidth at $\lambda = 1.35$~cm is $2.6^\prime$. The antenna sensitivity is 25~Jy/K. The signal was analyzed by a 2048-channel autocorrelation spectrometer with a resolution of 6.1~kHz (0.0822~km/s). All spectra were corrected for the absorption in the Earth's atmosphere.

Figures~1--7 present the results of our observations. The horizontal axis is the radial velocity with respect to the Local Standard of Rest in km/s, and the vertical axis is the flux density in janskys. Because of wide flux variations, the graphs are plotted in different scales of the vertical axis. In Figure~3 at the left spectra are clipped at some levels, and central parts of the spectra are shown at the right. This is done in order to show both faint emission features and the powerful flare feature. The double arrow shows the scale of the vertical axis. The epochs of the observations are indicated.

In Figure~6 the feature at 41.8~km/s is also clipped to show fainter features. In another scale this feature is shown completely in panel (14). Each maximum is labeled with the date of the observations: at the right for the ascending branch and at the left for the descending. We have also used the kindly supplied results obtained by Sergio Poppi [16] at the 64-meter radio telescope in Sardinia and by Simona Righini [17] on March 13, 2017, at the 32-meter radio telescope in Medicina. The data were processed by Sergio Poppi. Figure~7 presents the results; it also shows the spectrum obtained at RT-22 in Pushchino on March 22, 2017.

Figure~8 shows the evolution of individual components throughout our monitoring. Velocity variations are shown in Figure~8(a). The data are plotted with different symbols depending on the flux magnitude. Variations of the radial velocities of four main features are approximated with dashed straight lines. Figures~8(b) and (c) show the flux variability of individual features. For flux maxima the corresponding radial velocities are given. Secondary maxima of the main emission feature are marked with vertical arrows.

\section{Discussion}

We observed main H$_2$O maser emission of IRAS 18316$-$0602 in a narrow interval of radial velocities, from 39 to 44~km/s. Faint high-velocity emission was occasionally observed. This differs considerably from the structure of the spectra obtained in 1989 by Palla et~al. [11] and in 1994 by Jenness et~al. [2]. In their spectra the main maser emission was at a velocity of about 45~km/s with flux densities of 725 and 109~Jy. Meanwhile, there is much in common with the spectrum obtained by Kurtz and Hofner [18] in September 1995, in the first turn that the main emission was observed at 41.6~km/s.

From 2002 to 2016 we have observed three superflares and weaker flares that can be described as periods of high maser activity. The strongest flares took place in the velocity interval 40.5--42.5~km/s and were associated with the emission features whose points in the graph are linked with straight line \emph{2}. The maxima of the superflare emission tend to shift toward higher velocities (Figure~8(a)). This can be related to structural changes of the maser sources.

Thus, the most powerful emission was associated with feature (or cluster) \emph{3}. Most stable in time was the emission of feature \emph{3}.

We now consider the evolution of the maser emission flares in the chronological order.

\subsection{Superflare of 2002}

The beginning of our monitoring of IRAS 18316$-$0602 fell onto the descending branch of very high H$_2$O maser activity. In July 2002 the flux density of the main feature at 41~km/s was 3500~Jy. This means that at the flare maximum $F$ was $>3500$~Jy. During the evolution of this flare the linewidth did not depend on the flux density; it remained constant at 0.64~km/s. The line was well approximated with a gaussian, though it was slightly asymmetric: the right wing of the line is somewhat shallower than the left one.

During the flare of this feature the variability of the features at 39.4 and 43.3~km/s correlated with it. Then we observed sufficiently high activity of several features in the velocity interval 39.0--44.5~km/s with flux densities of up to 130~Jy. During 2004 we observed a feature at 40.6~km/s with a peak flux density of 340~Jy (May 2005). After that a period of low maser activity began.

\subsection{Superflare of 2010}

The evolution of this emission had a complicated character (see Figures 3 and 8). The superflare was preceded by a period of high maser activity. In the emission two features dominated. The emission of the fainter one was stable in the velocity (40.7~km/s), and the flux density did not exceed 370~Jy. The second feature (\emph{2b} in Figure~8) was much stronger than the first one, and during 2009 its flux density was varying within 550--1630~Jy. At the same time its radial velocity was slowly decreasing from 41.9 to 41.6~km/s. This feature may be considered as a precursor of the flare.

In three days (from January 26 to 29, 2010) the flux density at 41.6~km/s increased from 1640 to 7330~Jy. The analysis of the variation of all parameters of the superflare emission (flux density, radial velocity, width of the line and its shape) has shown that powerful emission consecutively appeared in two features with peak flux densities 19,060 and 6300~Jy. The time interval between the maxima was 5~months. The flux density peak of the former one (\emph{2b}) was drifting in the radial velocity from 41.9 to 40.8~km/s (the superflare precursor taken into account). At the epoch of the maximum of the emission of this feature the line was symmetric with a width at half maximum of 0.67~km/s; it was well approximated with a gaussian.

The velocity of the other feature (\emph{2c} in Figure~8 was 41.8~km/s, and it did not change appreciably. The observed velocity drift of the features could be related to the accelerated motion of maser condensations under the action of a shock wave.

We come to the conclusion that the most powerful flare of 2010 is associated with feature \emph{2b} (Figure~8), which was in the active state for about two years and was not associated with \emph{2c}. Thus the flares of 2010 and 2016 (this one discussed below) are associated with different features (maser condensations).

Since the end of 2014 emission features began to appear at $V_{\textrm{LSR}}>45$~km/s; they were mostly short-lived.

\subsection{Superflare of 2016}

This has been the strongest flare in this source. We detected it
at the ascending branch of its evolution. During a one-day time
interval the flux density increased by a factor of 1.5. The peak
flux density reached 46,000~Jy. The flare was short-lived. Its
duration at the 0.5 level was about one month. The linewidth and
flux density correlated: the line narrowed with growing flux,
and when the flux decreased the line broadened again. Figure~9
shows this dependence in the coordinates ($\ln\,F$), (${\Delta
V})^{-2}$, where $F$ is the peak flux density in janskys and
$\Delta V$ is the linewidth at half maximum in km/s. The
experimental data are plotted with circles. The graph is fitted
with a straight line. At the maximum activity the line is
strictly symmetric and it is ideally fitted with a gaussian. At
other epochs of observations (both at the ascending and
descending branches of the evolution) the flux density was below
15,000~Jy, and the line right wing was slightly shallower than
the left one. The symmetry of the line at the activity maximum
and its small width testify that the emission is related to a
single maser condensation. The slight asymmetry at other epochs
and a small line shift in radial velocity may occur if the
condensation is inhomogeneous. The same line shape was observed
in the flare of 2002. At that time the line velocity was
41.05~km/s, which is only by 0.75~km/s lower than in 2016.
Possibly this is the same feature whose radial-velocity drift
for 14~years was 0.75~km/s. Powerful superflares were observed
earlier in other sources, e.g., in Orion KL at a velocity of
8~km/s, see Matveenko [19], Matveenko et~al. [20]. According to
Garay et~al. [21], enhanced activity of the source persisted in
1979--1987. At that time the source flux density exceeded
$10^6$~Jy. With a difference in the distances to Orion KL
(500~pc) and IRAS 18316$-$0602 (3.3~kpc) superflares in these
two sources are comparable, together with similar linewidths
($\sim 0.6$~km/s). The difference is only in the activity
duration.

At the descending branch of the 2016 powerful flare in
IRAS\,18316$-$0602 we observed a considerably enhanced activity
of the maser source in a broad interval of radial velocities
(35--56~km/s). Flux densities of some features reached 500~Jy.
As in the case of the main feature, their flux densities were
varying very rapidly: during 1--2~days they changed by a factor of
1.5--2. No organized structures were noted. Probably the
activity was enhanced in individual features or clusters of
features with similar radial velocities within $\sim 2$~km/s.

Thus the flare had global character for the maser source in IRAS
18316$-$0602 and most likely was associated with a strong shock
wave from the central source. Nearly correlated variations of
features' flux densities and fast emission decrease can take
place in the case of a compact cluster of maser condensations
and of their small sizes.

The lack of VLA maps and the fact that the strong emission comes
only in a narrow  interval of radial velocities do not allow us to
reveal organized structures, which, as a rule, exist in the form
of extended filaments or chains. It is interesting that the
linewidth  of the strongest emission at all epochs of observations
including 1990 [11] was 0.60--0.67~km/s.

\section{Results}

We report the results of monitoring in the water vapor line at
$\lambda =1.35$~cm of the source IRAS\,18316$-$0602 associated
with a region of active star formation. The observations have been
carried out on the RT-22 radio telescope of the Pushchino
Observatory (Russia) in 2002--2017.

We observed three superflares in 2002, 2010, and 2016 with peak
flux densities 3400, 19,000, and 46,000 Jy, respectively. They
took place within a narrow interval of radial velocities
(40.5--42.5 km/s) and might be associated with a passage of a
strong shock wave. We have found correlation between flux density
variations and linewidth for the strongest flare of 2016
indicating that the maser was unsaturated.

The emission of three main groups of features was dominating. We
observed a small radial-velocity drift of this emission. In 2016
their velocities were sufficiently close, about 41, 42, and
43~km/s. Probably, they were localized nearly in the sky plane or
in a compact group.

Faint emission was occasionally observed at $V_{\textrm{LSR}}<37$
and $V_{\textrm{LSR}}>45$~km/s. Meanwhile, the superflare of 2016
was accompanied by rather intense emission (up to 500~Jy) in a
velocity interval of 35--56~km/s. No organized structures were
revealed. Most probably, the emission came from individual
features or a cluster of features with close radial velocities.

\section*{Acknowledgments}

This work was supported by the Russian Foundation for Basic
Research (project code 15-02-07676).The authors are grateful to
the staff of the Pushchino Radio Astronomy Observatory for the
great help with the observations, to Jan Brand, Sergio Poppi,
and Simona Righini for the results of H$_2$O observations at the
64-m and 32-m radio telescopes in Sardinia and Medicina.

\begin{figure}
\centering\leavevmode \epsfxsize=0.87\textwidth
\epsfbox{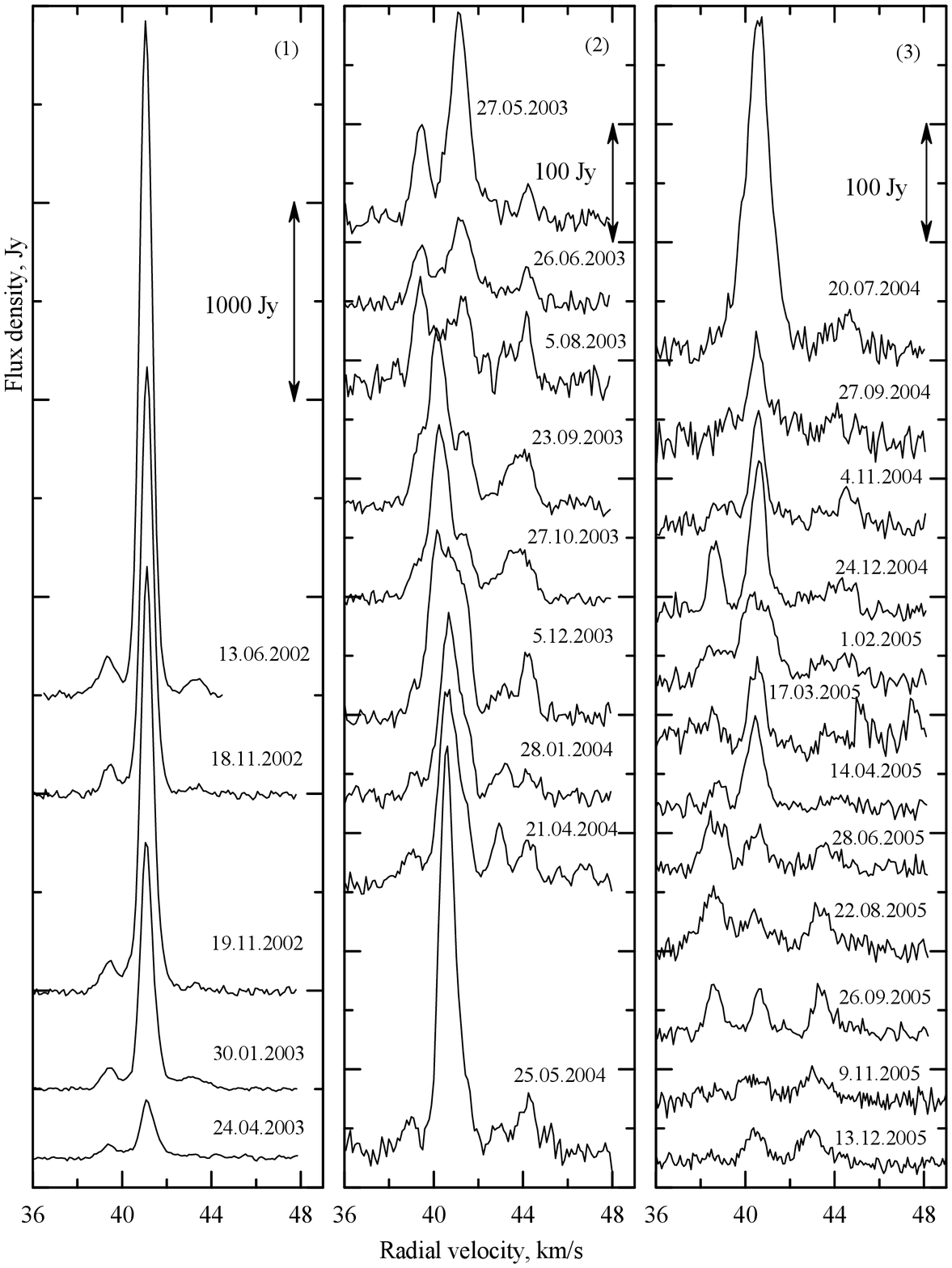} \caption{\small H$_2$O $\lambda
=1.35$~cm maser emission spectra of IRAS\,18316$-$0602 in
2002--2005. Double vertical arrows show the scale in janskys. The
epochs of the observations are given.} \label{fig1}
\end{figure}

\newpage

\begin{figure}
\centering\leavevmode \epsfxsize=.9\textwidth
\epsfbox{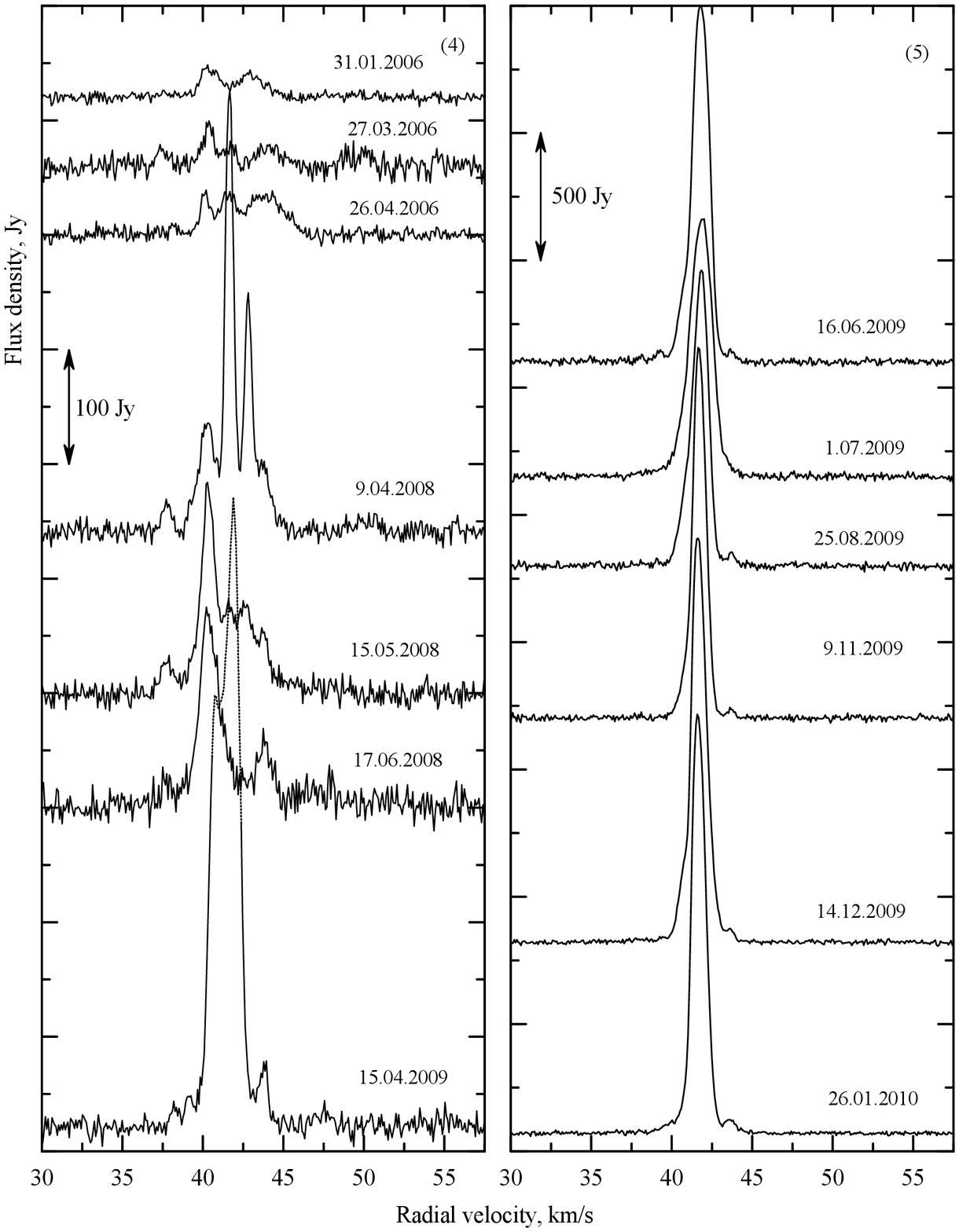} \caption{\small H$_2$O maser emission
spectra of IRAS\,18316$-$0602 in 2006--2010.}
\label{fig2}
\end{figure}

\newpage

\begin{figure}
\centering\leavevmode \epsfxsize=.82\textwidth
\epsfbox{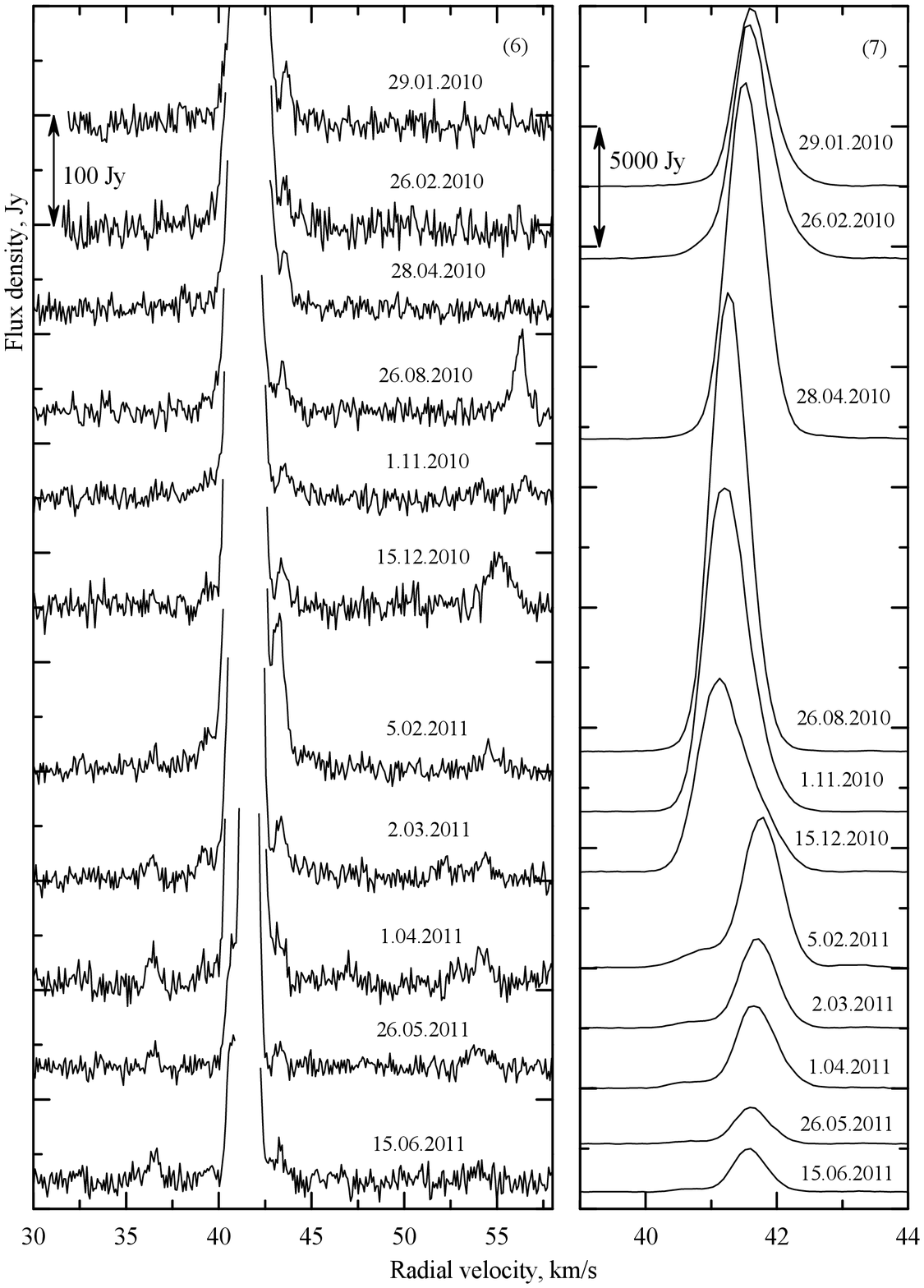} \caption{\small H$_2$O maser emission
spectra of IRAS\,18316$-$0602 in 2010--2011.}
\label{fig3}
\end{figure}

\newpage

\begin{figure}
\centering\leavevmode \epsfxsize=.9\textwidth
\epsfbox{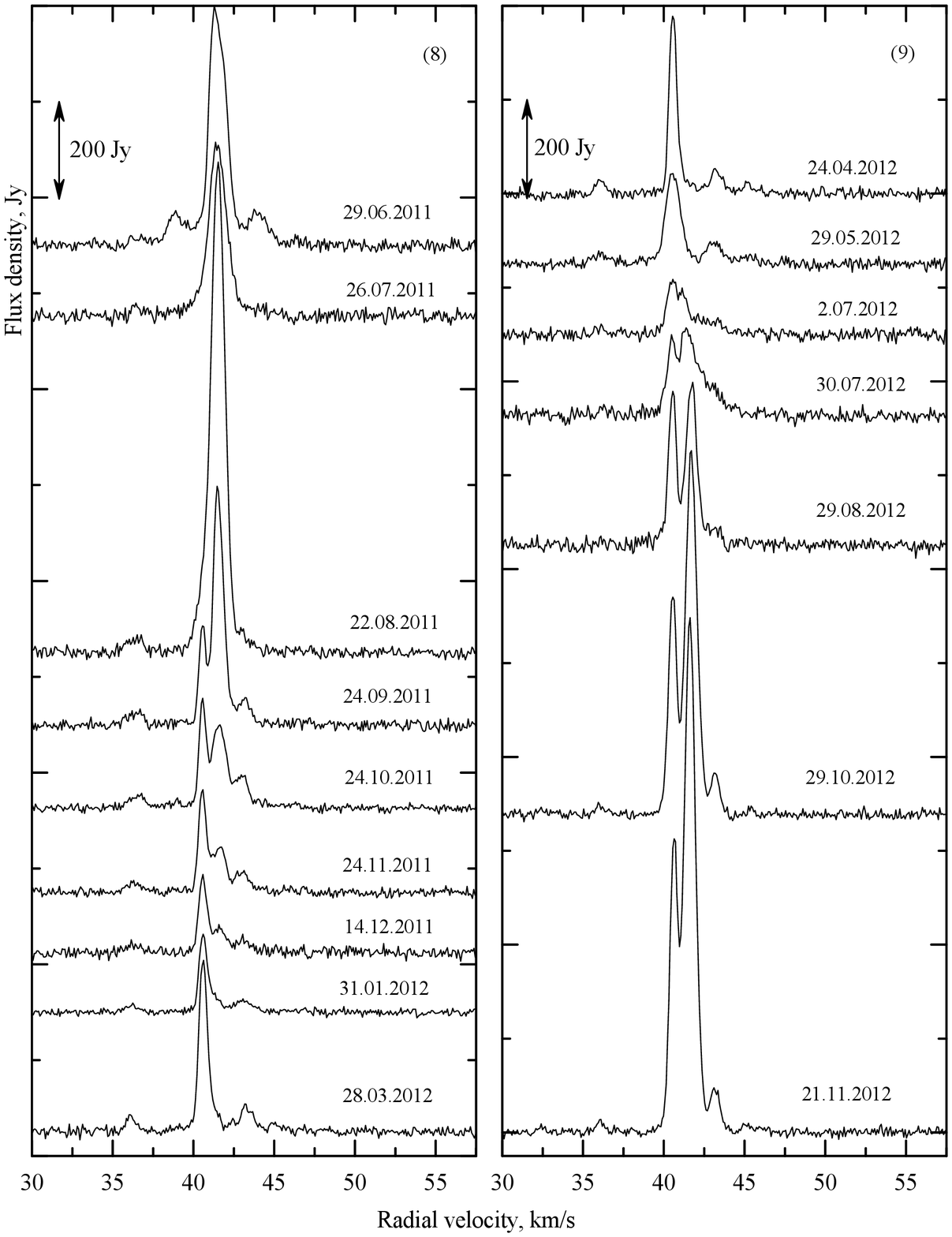} \caption{\small H$_2$O maser emission
spectra of IRAS\,18316$-$0602 in 2011--2012.}
\label{fig4}
\end{figure}

\newpage

\begin{figure}
\centering\leavevmode \epsfxsize=.9\textwidth
\epsfbox{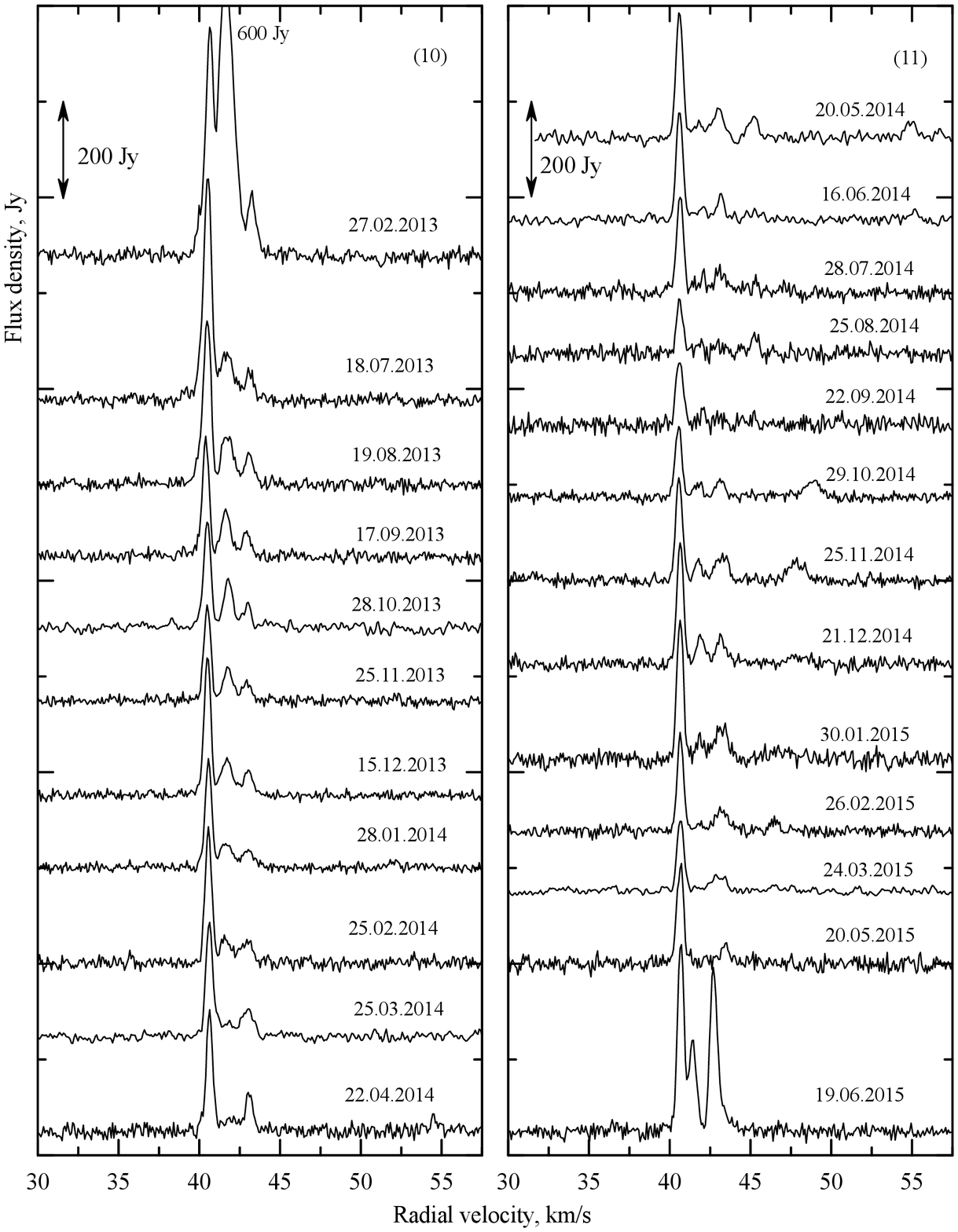} \caption{\small H$_2$O maser emission
spectra of IRAS\,18316$-$0602 in 2013--2015.} \label{fig5}
\end{figure}

\newpage

\begin{figure}
\centering\leavevmode \epsfxsize=.9\textwidth
\epsfbox{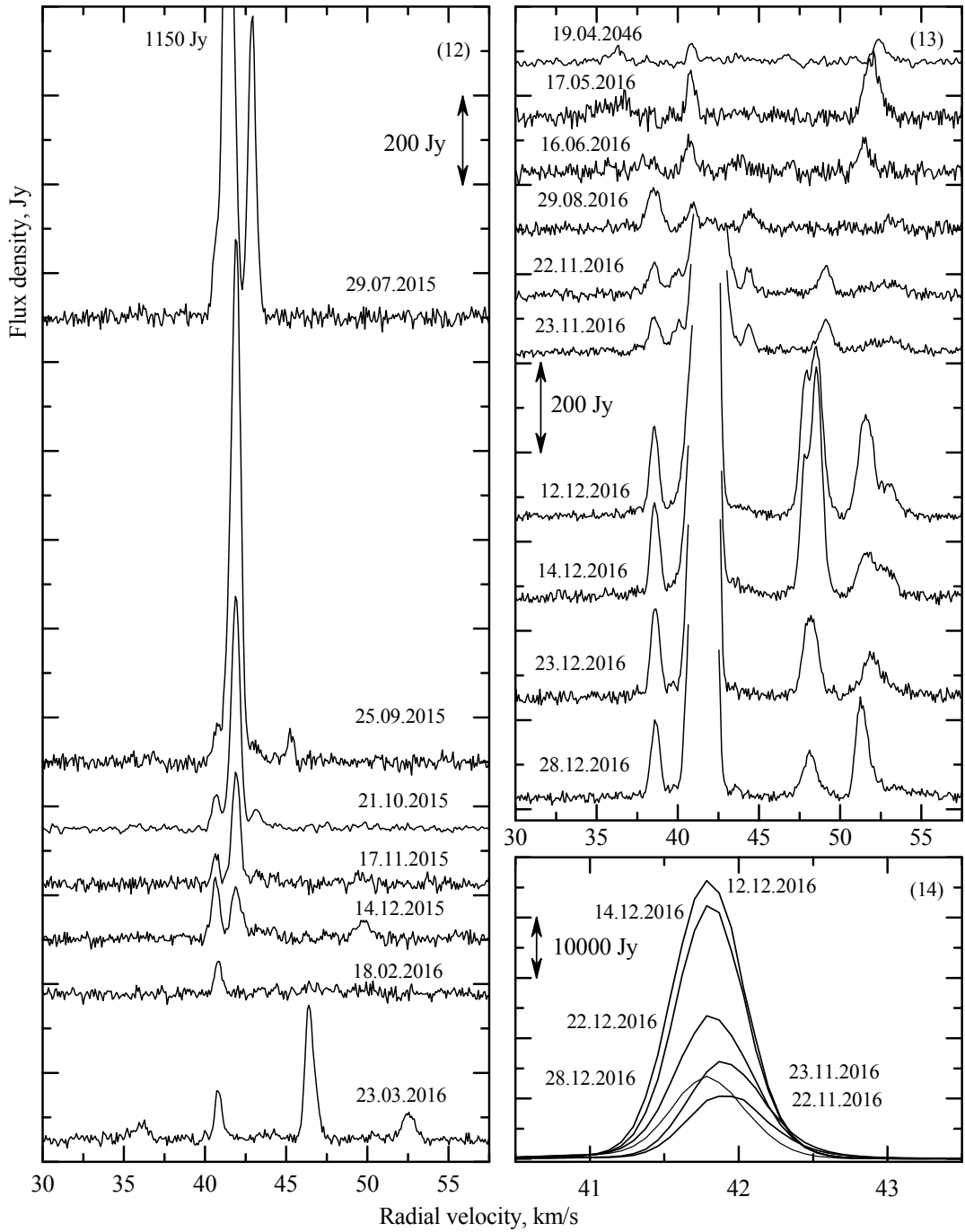} \caption{\small H$_2$O maser emission
spectra of IRAS\,18316$-$0602 in 2015--2016. The central feature
is shown in full in panel (14). Each maximum is labeled with the
date of the observations; left: for the ascending branch, right:
for the descending.} \label{fig6}
\end{figure}

\newpage

\begin{figure}
\centering\leavevmode \epsfxsize=.6\textwidth
\epsfbox{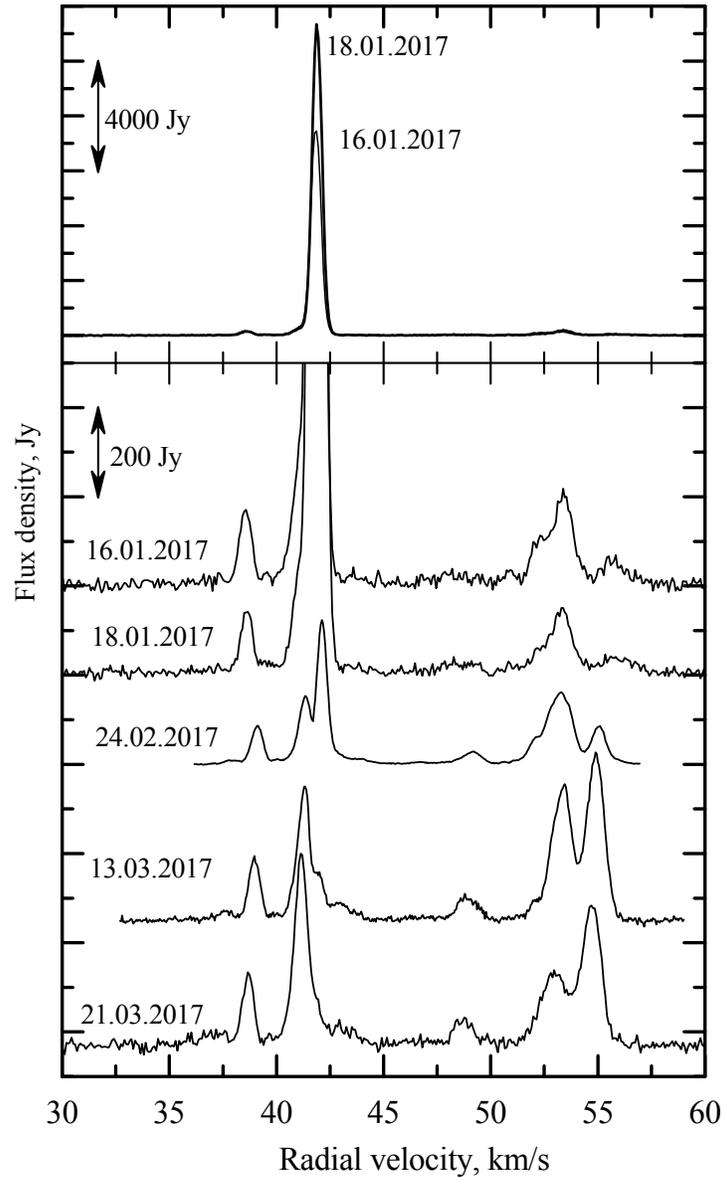} \caption{\small H$_2$O maser emission
spectra of IRAS\,18316$-$0602 in 2017.} \label{fig7}
\end{figure}

\newpage

\begin{figure}
\centering\leavevmode \epsfxsize=0.8\textwidth
\epsfbox{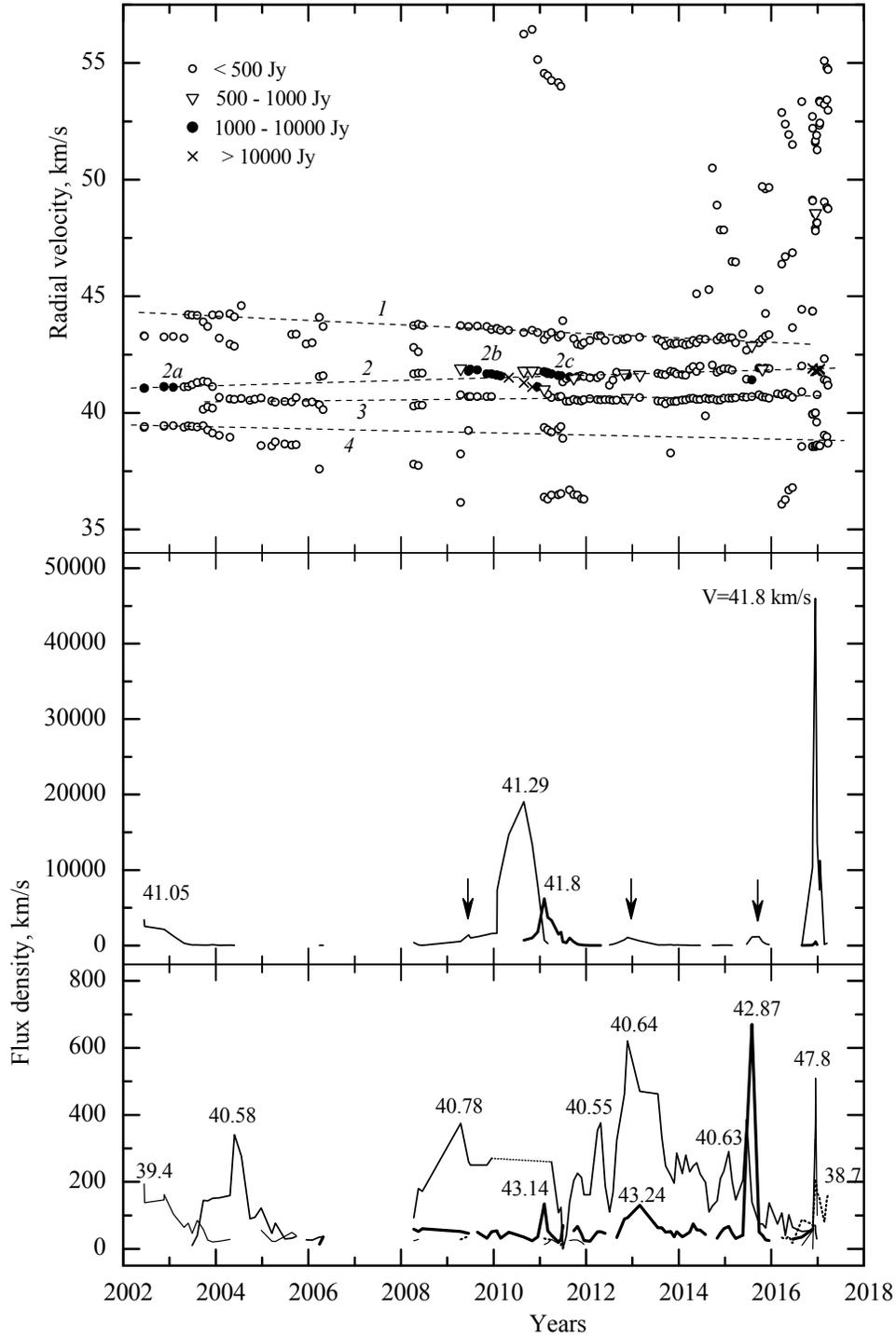} \caption{\small Variability of the
radial velocity (a) and flux density (b, c) of the main spectral
features. Radial-velocity variations of three main features are
approximated with straight (dashed) lines and are numbered. Flux
density maxima are labeled with the corresponding radial
velocities. Fainter maxima of the main emission feature are marked
with vertical arrows.} \label{fig8}
\end{figure}

\newpage

\begin{figure}
\centering\leavevmode \epsfxsize=.5\textwidth
\epsfbox{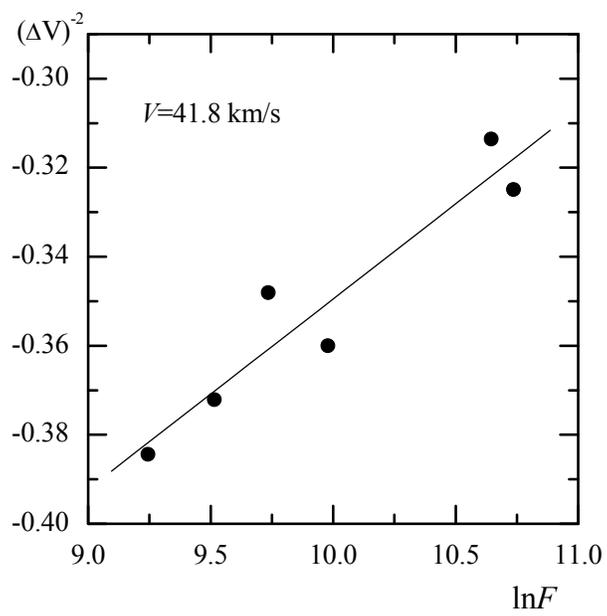} \caption{\small Connection between
variations of the flux density and linewidth for the 41.8-km/s
emission feature.} \label{fig9}
\end{figure}

\end{document}